\documentclass[12pt]{iopart}
 
\newcommand{\zp}{Z^\prime}
\newcommand{\numue}{\nu_\mu e}
\newcommand{\nuee}{\nu_e e}
\newcommand{\nuelle}{\nu_\ell e}
\newcommand{\bnumue}{\bar\nu_\mu e}
\newcommand{\sintthw}{\sin^2\theta_W}
\newcommand{\sinfthw}{\sin^4\theta_W}
\newcommand{\sutwo}{SU(2)$_L \times{}$U(1)$_Y$}
\newcommand{\be}{\begin{equation}}
\newcommand{\ee}{\end{equation}}
\newcommand{\bea}{\begin{eqnarray}}
\newcommand{\eea}{\end{eqnarray}}
\newcommand{\noi}{\noindent}
\newcommand{\eps}{\epsilon}
\newcommand{\overms}{\overline{MS}}
 
\def\parenbar#1{{\null\!                        
   \mathop{\smash#1}\limits
   ^{\hbox{\tiny(--)}}
   \!\null}}                                    
 
\def\nnbar{\parenbar{\nu}}

\newcommand{\ltsim}{\mathrel{\lower4pt\hbox{$\sim$}}
\hskip-12.5pt\raise1.6pt\hbox{$<$}\;}
 
\newcommand{\gtsim}{\mathrel{\lower4pt\hbox{$\sim$}}
\hskip-11.5pt\raise1.6pt\hbox{$>$}\;}

\begin{document}
 
\title{Neutrino-Electron Scattering Theory}
 
\author{William J. Marciano and Zohreh Parsa}
 
\address{Brookhaven National Laboratory,
Upton, New York\ \ 11973}
 
\begin{abstract}
Standard Model predictions for neutrino-electron scattering
cross-sections, including effects of electroweak radiative corrections,
are reviewed.  The sensitivity of those quantities to neutrino dipole
moments, $\zp$ bosons, and dynamical symmetry breaking is described.
Neutrino indices of refraction in matter
are also discussed. A perspective on future initiatives with intense
neutrino sources, such as from stopped pion decays at a neutron 
spallation source, superbeams or  neutrino factories, is given.
\end{abstract}
\vfill
 
{\footnotesize \noi This manuscript has been authored under contract
number DE-AC02-98CH10886
with the U.S. Department of Energy.  Accordingly, the U.S.
Government retains a non-exclusive, royalty-free license to publish
or reproduce the published form of this contribution,
or allow others to do so, for U.S. Government purposes.}
 
\maketitle

\section{Introduction}
Neutrino-electron scattering cross-sections are extremely small and
correspondingly very difficult to measure. Nevertheless, because of
heroic experimental efforts, they have played a crucial role in
confirming the SU(2)$_L \times{}$U(1)$_Y$ structure of the Standard
Model (SM) and in helping to unravel subtle properties of neutrinos. In
particular, the initial observation of $\bar\nu_\mu e\to \bar\nu_\mu e$
scattering at CERN \cite{one} confirmed the existence of weak neutral
currents. Subsequently, higher statistics studies \cite{two} of
$\numue$ and $\bnumue$ scattering provided a clean (purely leptonic)
determination of the weak mixing angle, $\sintthw$. On a separate
front, low energy studies of $\nu e\to \nu e$ solar neutrino scattering
by the Super $K$ Collaboration \cite{three} helped unveil the nature of
neutrino mixing and oscillations by exploiting specific SM differences
between $\nuee$ and $\nuelle$, ($\ell=\mu$ or $\tau$), scattering
cross-sections.
 
Agreement between measured neutrino-electron and antineutrino-electron
cross-sections and SM expectations has also been used to constrain
physics beyond the SM, (i.e.\ ``New Physics'') effects. Bounds on
neutrino dipole moments (magnetic, electric and transition) at about
the $10^{-9}$--$10^{-10} e/2 m_e$ level have been given \cite{marc}.
Constraints on
$\zp$ bosons, excitations of extra dimensions, dynamical symmetry
breaking etc.\ can also be extracted. However, currently, they are
usually not as stringent as other precision electroweak tests. In the
future, more precise studies of  $\nu e$ scattering could be made
competitive by utilizing high intensity neutrino sources such as
stopped pion decays at neutron spallation sources,
superbeams or neutrino factories. The last one would allow for intense
high energy $\nu_e$, $\bar\nu_\mu$, $\bar\nu_e$ and $\nu_\mu$ beams
derived from $\mu^\pm$ decays in storage rings. The statistical figure
of merit for cross-section measurements with such beams grows as
$E^3_\mu$; so, high energy is of prime importance. If constructed, such
facilities would play a valuable role in the next generation of ``New
Physics'' probes, with $\nu e$ scattering providing  a small part of
their
extensive physics programs \cite{mang}.
 
Of course, to fully utilize precision $\nu e$ cross-section
measurements, electroweak radiative corrections must be included. Such
effects have been fully computed at the one loop level \cite{saran} and
if required,
two loop calculations (albeit tedious and demanding) could be carried
out. Theoretical efforts to push those calculations much further would
require motivation from new experiments.
 
In this overview of $\nu e$ scattering theory, our plan is as follows.
First, we give in section~2 the SM predictions for such cross-sections.
Then in section~3, the order $\alpha$ (one loop + bremsstrahlung)
electroweak radiative corrections to those reactions are presented.
Effects of ``New Physics'', including neutrino dipole moments, $\zp$
bosons and dynamical symmetry breaking are discussed in section~4. A
somewhat different, but related topic, neutrino indices of refraction
in matter, is reviewed in section~5. That interesting phenomenon which
finds application in terrestrial, solar and supernova studies stems
from neutrino forward scattering amplitudes with the constituents of
matter (which include electrons). Hence, we deem that  subject
appropriate for this article. Finally, in section~6, we give a
perspective for future neutrino-electron studies that would be made
possible by intense neutrino sources such as stopped pion decays at neutron
spallation sources,  superbeams or neutrino
factories.
 
\section{Tree Level Cross-Sections}
Within the SM \sutwo\ framework of electroweak interactions, a variety
of neutrino and antineutrino scattering cross-sections are possible.
Here, we divide them into 3 categories. The first set, although
kinematically suppressed, is conceptually simple. It corresponds to
pure $W$ exchange (charged current interactions). In the $t$-channel,
one can have
 
\be
\nu_\ell+e\to\ell+\nu_e \qquad (\ell=\mu{\rm~or~}\tau) \label{eqone}
\ee
 
\noi while in the $s$-channel
 
\be
\bar\nu_e +e\to\ell+\bar\nu_\ell \qquad (\ell=\mu{\rm~or~}\tau)
\label{eqtwo}
\ee
 
\noi These are sometimes referred to as inverse muon (or tau) decays.
The threshold (for electrons at rest)
 
\be
E_\nu\ge \frac{m^2_\ell - m^2_e}{2m_e} \label{eqthree}
\ee
 
\noi is quite high for $\ell=\mu$, i.e.\ $E_\nu\gtsim 10.8$ GeV and
essentially unaccessible for $\ell=\tau$, $E_\nu\gtsim 3$ TeV\null.
Nevertheless, for generality we leave $\ell$ arbitrary. Note also that
for the $s$-channel in eq~(\ref{eqtwo}) semileptonic reactions
$\bar\nu_e+e\to d+\bar u$, $s+\bar u$, $d+\bar c$ etc.\ are also
possible. However, those possibilities will not be discussed in this
article. Instead, we focus only  on leptonic reactions.
 
Pure $Z$ exchange in the $t$-channel (weak neutral currents) give rise
to a second more easily accessible set of reactions
 
\be
\nnbar_\ell + e \to \nnbar_\ell +e \qquad (\ell=\mu{\rm~or~}\tau)
\label{eqfour}
\ee
 
\noi where the short-hand notation $\nnbar$ stands for $\nu$ or
$\bar \nu$.
 
Finally, the third possibility
 
\be
\nnbar_e +e\to \nnbar_e+e \label{eqfive}
\ee
 
\noi proceeds through a combination of $W$ and $Z$ exchange amplitudes.
 
Before reviewing the tree level predictions for the above processes, we
state our simplifying assumptions. Neutrino masses and mixing are
neglected (except indirectly in section~5 when neutrino indices of
refraction are discussed). We assume $|q^2|\ll m^2_W$ or $m^2_Z$; so,
propagator effects can be ignored and effective four-fermion amplitudes
employed. Because the electron target is at rest, that is a very good
approximation.  All neutrino scattering amplitudes are normalized in
terms of the Fermi constant, $G_\mu$, obtained from the muon decay rate
(lifetime) \cite{marctwo}
 
\be
G_\mu = \frac{g^2}{4\sqrt{2}m^2_W} = 1.16637 (1)\times10^{-5} {\rm
~GeV}^{-2} \label{eqsix}
\ee
 
\noi That quantity is very accurately determined and will prove useful
in section~3 when electroweak radiative corrections are considered.
 
With those assumptions, charged current reactions result from the
effective tree level amplitude
 
\be
M_{cc} = - i \frac{G_\mu}{\sqrt{2}} \bar u_\ell \gamma^\alpha
(1-\gamma_5) u_{\nu_\ell} \bar u_{\nu_e} \gamma_\alpha (1-\gamma_5) u_e
\label{eqseven}
\ee
 
\noi where the $u_f$ are 4 component spinors corresponding to their
subscript fermions. To obtain the differential cross-sections, one
squares the amplitude, $|M_{cc}|^2$, averages over the initial electron
polarizations, sums over final state polarizations and integrates over
the unobserved final state neutrino momentum. In that way, one finds in
the electron rest frame (the lab system) for $\ell=\mu$ or $\tau$
 
\be
\frac{d\sigma(\nu_\ell e\to\ell\nu_e)}{dy} = \frac{G^2_\mu}{\pi} (2m_e
E_\nu - (m^2_\ell-m^2_e)) \label{eqeight}
\ee
 
\be
\frac{d\sigma(\bar\nu_e e\to\ell\bar\nu_\ell)}{dy} =
\frac{G^2_\mu}{\pi} (2m_e E_\nu (1-y)^2 - (m^2_\ell - m^2_e) (1-y))
\label{eqnine}
\ee
 
\noi where $E_\nu$ is the initial state neutrino energy and
 
\be
y=\frac{E_\ell - \frac{m^2_\ell+m^2_e}{2m_e}}{E_\nu} \label{eqten}
\ee
 
\noi with $E_\ell$ the final state charged lepton energy. The range of
$y$ is
 
\be
0 \le y \le y_{\rm max} = 1-\frac{m^2_\ell}{2m_e E_\nu+m^2_e}
\label{eqeleven}
\ee
 
\noi Note that at threshold, $E_\nu=\frac{m^2_\ell-m^2_e}{2m_e}$, the
range of integration collapses to zero.
 
Integrating over $y$, one finds in the very high energy (extreme
relativistic) limit $E_\nu \gg \frac{m^2_\ell-m^2_e}{2m_e}$
 
\bea
\sigma(\nu_\ell e\to\ell\nu_e) \simeq 3\sigma (\bar\nu_e
e\to\ell\bar\nu_\ell) & \simeq & \frac{2G^2_\mu m_e E_\nu}{\pi}
\nonumber \\
& \simeq & 1.5\times10^{-41} (E_\nu/{\rm GeV}){\rm~cm}^2
\label{eqtwelve}
\eea
 
The pure neutral current reactions in eq~(\ref{eqfour}) can be analyzed
in the same way. Their kinematics is simpler since $m_\ell\to m_e$, but
a slight complication is a combination of left and right-handed
electron couplings in the effective amplitude ($\ell=\mu$ or $\tau$)
 
\be
M_{\rm NC} = i\frac{G_\mu}{\sqrt{2}} \bar u_{\nu_\ell} \gamma^\alpha
(1-\gamma_5) u_{\nu_\ell} [ \eps_- \bar u_e \gamma_\alpha (1-\gamma_5)
u_e + \eps_+ \bar u_e\gamma_\alpha (1+\gamma_5) u_e] \label{eqthirteen}
\ee
 
\noi where at the tree level \cite{thooft}
 
\bea
\eps_- & = & \frac12 -\sintthw \nonumber \\
\eps_+ & = & -\sintthw \label{eqfourteen}
\eea
 
In terms of
 
\be
y=\frac{E^\prime_e-m_e}{E_\nu}, \qquad 0\le y\le y_{\rm max} =
\frac{1}{1+m_e/2E_\nu} \label{eqfifteen}
\ee
 
\noi with $E^\prime_e$ the final state electron energy, the
differential cross-sections for $\nnbar_\ell e\to \nnbar_\ell e$ are
given by (for $\ell=\mu$ or $\tau$)
 
\bea
\frac{d\sigma(\nu_\ell e\to\nu_\ell e)}{dy} & = & \frac{2G^2_\mu m_e
E_\nu}{\pi} [\eps^2_- + \eps^2_+(1-y)^2 - \eps_-\eps_+
\frac{m_e}{E_\nu} y] \label{eqsixteen} \\
\frac{d\sigma(\bar\nu_\ell e\to\bar\nu_\ell e)}{dy} & = &
\frac{2G^2_\mu m_e E_\nu}{\pi} [\eps^2_+ + \eps^2_- (1-y)^2
-\eps_-\eps_+ \frac{m_e}{E_\nu}y] \label{eqseventeen}
\eea
 
\noi where a small $\eps_-\eps_+$ interference term has been retained
for low energy applications. Note that $\sigma(\nu_\ell e\to\nu_\ell
e)$ and $\sigma(\bar\nu_\ell e\to\bar\nu_\ell e)$ are related by
$\eps_- \leftrightarrow \eps_+$ interchange \cite{saran}.
 
Neglecting terms of relative order $m_e/E_\nu$, one finds for the
integrated cross-sections
 
\bea
\sigma (\nu_\ell e \to \nu_\ell e) & = & \frac{G^2_\mu m_e E_\nu}{2\pi}
[1-4\sin^2\theta_W + \frac{16}{3} \sin^4\theta_W] \label{eqeighteen} \\
\sigma (\bar\nu_\ell e \to \bar\nu_\ell e) & = & \frac{G^2_\mu m_e
E_\nu}{2\pi}
[\frac13-\frac43\sin^2\theta_W + \frac{16}{3} \sin^4\theta_W]
\label{eqnineteen}
\eea
 
\noi For $\sin^2\theta_W \simeq 0.23$, those cross-sections are very
small $\sim 10^{-42} (E_\nu/{\rm GeV}) {\rm cm}^2$. Nevertheless, they
have been rather well measured (for $\ell=\mu$) \cite{two}, yielding
$\sin^2\theta_W$ to about $\pm3.5$\%. At that level, electroweak
radiative corrections become important and must be applied in any
serious study. Indeed, they must be considered just to define
$\sintthw$ in a meaningful way. These corrections will be discussed in
section~3.
 
The final cross-sections to be considered are those in
eq~(\ref{eqfive}) that result from combined $W$ and $Z$ boson exchange.
They are obtained at tree level from eqs.~(\ref{eqsixteen}) and
(\ref{eqseventeen}) under the replacements (for $\nnbar_\ell \to
\nnbar_e)$ \cite{saran}
 
\bea
\eps_- \to \eps^\prime_- & = & \eps_- -1 = -\frac12 -\sintthw \nonumber
\\
\eps_+ \to \eps^\prime_+ & = & \eps_+ = -\sintthw \label{eqtwenty}
\eea
 
\noi Again ignoring $m_e/ E_\nu$ effects, one finds.
 
\bea
\sigma (\nu_\e e \to \nu_\e e) & = & \frac{G^2_\mu m_e E_\nu}{2\pi}
[1+4\sin^2\theta_W + \frac{16}{3} \sin^4\theta_W] \nonumber \\
\sigma (\bar\nu_\e e \to \bar\nu_\e e) & = & \frac{G^2_\mu m_e
E_\nu}{2\pi}
[\frac13+\frac43\sin^2\theta_W + \frac{16}{3} \sin^4\theta_W]
\label{eqtwentyone}
\eea
 
\noi Those cross-sections are roughly a factor of 7 and 3 respectively
larger than those in eqs.~(\ref{eqeighteen}) and (\ref{eqnineteen}).
The significant difference between $\sigma(\nu_e e\to\nu_e e)$ and
$\sigma(\nu_\ell e\to \nu_\ell e)$, $\ell=\mu$ or $\tau$ has played a
key role in sorting out what fraction of solar neutrinos reach the
earth as $\nu_e$ and what fraction arrive as $\nu_\ell$ ($\ell=\mu$ or
$\tau$). That information is important for unfolding neutrino mixing
and oscillations \cite{three}.
 
In table~\ref{tabone}, we summarize the tree level predictions and
relative sizes for the scattering cross-sections discussed in
this section.  Precise measurements of those cross-sections are generally
limited by systematic uncertainties in the neutrino flux and spectrum.
To help overcome that limitation, various ratios of cross sections are
often discussed. Two cases  considered for low energy studies are
\cite{two}

\be
R_1 \equiv \frac{\sigma(\nu_\mu e\to \nu_\mu e)}{\sigma(\bar\nu_\mu
e\to\bar\nu_\mu e)} \label{eqtwentytwo}
\ee
 
\noi and
 
\be
R_2 = \frac{\sigma(\nu_\mu e\to\nu_\mu e)}{\sigma(\nu_e e\to \nu_e e) +
\sigma(\bar\nu_\mu e\to \bar\nu_\mu e)} \label{eqtwentythree}
\ee
 
\noi The latter, $R_2$, would use low energy neutrinos from $\pi^+$
and $\mu^+$ decays in the chain
 
\be \pi^+\to \mu^+\nu_\mu \qquad , \qquad \mu^+\to e^+\nu_e \bar\nu_\mu
\label{eqtwentyfour}
\ee
 
\noi to normalize the flux in that ratio. It was proposed \cite{white}
for an
experiment at LAMPF, but not carried out. It would be useful for
neutrino physics at an intense neutron spallation facility where many
$\pi^+$ are produced. If flux normalizations can be controlled, then
one expects at tree level
 
\bea
R_1 & = & \frac{3-12\sintthw +16\sin^4\theta_W}{1-4\sintthw
+16\sin^4\theta_W} \label{eqtwentyfive} \\
R_2 & = & \frac{3-12\sintthw +16\sin^4\theta_W}{4+8\sintthw
+32\sin^4\theta_W} \label{eqtwentysix}
\eea
 
\begin{table}
\caption{Relative size of various tree level $\nnbar e$ cross-sections
in units of $\frac{G^2_\mu m_e E_\nu}{2\pi}$ for the limit $E_\nu \gg
m^2_\mu /2m_e$ but $-q^2\ll m^2_W$. \label{tabone}}
\begin{center}
\begin{tabular}{lcc}
Reaction & $\sigma/(G^2_\mu m_e E_\nu/2\pi)$ & Relative Size
($\sintthw=0.23$) \\ \\
$\nu_\mu e\to \mu^-\nu_e$ & 4 & 4 \\
$\bar\nu_e e\to \mu^-\bar\nu_\mu$ & 4/3 & 4/3 \\
$\nu_\mu e \to \nu_\mu e$ & $1-4\sintthw + \frac{16}{3}\sinfthw$ & 0.362
\\
$\bar\nu_\mu e\to \bar\nu_\mu e$ &
$\frac13-\frac43\sintthw+\frac{16}{3}\sinfthw$ & 0.309 \\
$\nu_e e\to \nu_e e$ & $1+4\sintthw+\frac{16}{3} \sinfthw$ & 2.2 \\
$\bar\nu_e e\to \bar\nu_e e$ &
$\frac13+\frac43\sintthw+\frac{16}{3}\sinfthw$ & 0.922
\end{tabular}
\end{center}
\end{table}

Neutrino factories \cite{mang} offer the best solution to flux
normalization. The
decay possibilities $\mu^-\to e^-\bar\nu_e\nu_\mu$ and $\mu^+\to
e^+\nu_e\bar\nu_\mu$ from a long straight section at a muon storage
ring would have very well specified neutrino energy spectra. One
possibility
would be to run in both modes and measure (after weighting for the
different spectra)
 
\be
R_3 = \frac{\sigma(\bar\nu_ee\to\bar\nu_ee) + \sigma(\nu_\mu
e\to\nu_\mu e)}{\sigma (\nu_ee\to\nu_ee) + \sigma(\bar\nu_\mu
e\to\bar\nu_\mu e)} \label{eqtwentyseven}
\ee
 
\noi which is predicted at tree level to be
 
\be
R_3 = \frac{1-2\sintthw +8\sinfthw}{1+ 2\sintthw+8 \sinfthw}
\label{eqtwentyeight}
\ee
 
\noi At a very high energy muon storage ring, e.g.\ $E_\mu\simeq50$
GeV, where the average neutrino energies are $\sim40$ GeV, one can use
the reactions $\bar\nu_ee\to\mu^-\bar\nu_\mu$ and $\nu_\mu
e\to\mu^-\nu_\mu$ from the $\mu^-$ decay neutrinos to normalize the flux.
That possibility corresponds to measuring (after accounting for different 
spectra)
 
\be
R_4 \equiv \frac{\sigma(\bar\nu_ee\to\bar\nu_ee) + \sigma(\nu_\mu
e\to\nu_\mu e)}{\sigma(\bar\nu_ee\to\mu^-\bar\nu_\mu) + \sigma(\nu_\mu
e\to \mu\nu_\mu)} \label{eqtwentynine}
\ee
 
\noi which is predicted at tree level to be (using the high energy
limits in eq~(\ref{eqtwelve}))
 
\be
R_4 = \frac14 (1-2\sintthw + 8\sinfthw) \label{eqthirty}
\ee
 
Of course, to properly utilize such quantities either to extract
$\sintthw$ with high precision or to search for signs of ``New
Physics'' requires inclusion of electroweak radiative corrections, a
subject we now address.
 
\section{Electroweak Radiative Corrections}
The full $\cal{O}(\alpha)$ electroweak radiative corrections to
neutrino-electron scattering in the Standard Model were computed
\cite{saran}
in 1983. Here, we provide a brief summary of the results, generally
making the same simplifying kinematic assumptions as in section~2.
Also, we now ignore effects of relative order $\alpha m_e/ E_\nu$
(left-right interference). A detailed study of the latter for low
energy neutrinos can be found in ref.~\cite{pass}.
 
$\cal{O}(\alpha)$ corrections to $\nnbar e$ scattering cross-sections
include the full one loop electroweak corrections of the SM as well as
photon
bremsstrahlung effects. The loops contain short-distance ultraviolet
divergences which are canceled by renormalization counterterms induced
by replacing bare couplings and masses with renormalized quantities.
The usual prescription \cite{marcthree} involves replacing
$g^2_0/4\sqrt{2} m^{0^2}_W$
by $G_\mu$ and $\sin^2\theta^0_W$ by some appropriately defined
renormalized weak mixing angle. Here, we employ a modified minimal
subtraction $(\overms)$ definition \cite{marcfour}
 
\be
\sintthw(\mu)_{\overms} \label{eqthirtyone}
\ee
 
\noi where $\mu$ is the 't Hooft unit of mass in dimensional
regularization. We later specialize to $\mu=m_Z$.  When the $\nnbar e$
cross-sections are expressed in
terms of $G_\mu$ and $\sintthw(\mu)$, the radiative corrections become
finite and calculable. Of course, the results can be easily translated
into other schemes, for example the on-shell definition \cite{sirlin}
$\sintthw\equiv 1-m^2_W/m^2_Z$.
 
Specializing to the extreme relativistic limit where terms of relative
order $m_e/E_\nu$ and $m^2_e/q^2$ can be ignored, one finds that the
bulk of the loop corrections to the charged current reactions in
eqs.~(\ref{eqone}) and (\ref{eqtwo}) are absorbed in $G_\mu$. The
remaining loop effects plus bremsstrahlung modify the differential
cross-sections in eqs.~(\ref{eqeight}) and (\ref{eqnine}) by the
overall factors \cite{saran}
 
\bea
&&1+\frac{\alpha}{\pi} f_-(y) \nonumber \\
f_-(y) & = & -\frac23\ell n \frac{2E_\nu}{m_e} + (\ell n (1-y) -
\frac12\ell n y + \frac{y}{2}+\frac14) \ell n \left(\frac{2m_e
E_\nu}{m^2_\ell}\right) \label{eqthirtytwo} \\
&& + \frac12 \left[L(y) + \frac{\pi^2}{6}\right] -\frac12 \ell n^2
\left(\frac{1-y}{y}\right) +y\ell n y - \left( \frac{23}{12} +
\frac{y}{2}\right) \ell n (1-y) \nonumber \\
&& -\frac{47}{36} - \frac{11}{12} y + \frac{y^2}{24} \nonumber
\eea
 
\noi for $d\sigma (\nu_\ell e\to\ell\nu_e)/dy$ and
 
\bea
&& 1+\frac{\alpha}{\pi} f_+(y) \nonumber \\
(1-y)^2f_+(y) &=& -\frac23 (1-y)^2 \ell n \frac{2E_\nu}{m_e} +
\left[\left(y(1-y)-\frac12\right) \ell n y + (1-y)^2 \ell n
(1-y)\right. \nonumber \\
&& \left.- \frac{1-y}{2}\right] \ell n \frac{2m_e E_\nu}{m^2_\ell}
\label{eqthirtythree} \\
&& + \left(y (1-y) -\frac12\right) \left(\ell n^2
y-\frac{\pi^2}{6}-L(y)\right) \nonumber \\
&&  + (1-y)^2 \ell n (1-y)\left[\ell n y -\frac12 \ell n (1-y)\right]
\nonumber \\
&& +(\ell n y)\left(-\frac34+\frac{y}{2}+y^2\right) +\frac13 (1-y) \ell
n (1-y) \left(-\frac72+5y\right)\nonumber \\
&& -\frac{1-y}{72} (31-49y) \nonumber
\eea
 
\noi for $d\sigma(\bar\nu_ee\to \ell \bar\nu_\ell)$, where
 
\be
L(y) = \int^y_0 dt \frac{\ell n(1-t)}{t} \label{eqthirtyfour}
\ee
 
\noi Integrating those expressions over $y$ (assuming $E_\nu\gg
m^2_\mu/2m_e$), one finds
 
\bea
\sigma(\nu_\ell e\to\ell\nu_e) \simeq \frac{2G^2_\mu m_e E_\nu}{\pi}
\left(1+\frac{\alpha}{\pi} F_-\right) \nonumber \\
F_- = -\frac23 \ell n \frac{2E_\nu}{m_e} -\frac16 \left( \pi^2 -
\frac{19}{4}\right) \label{eqthirtyfive}
\eea
 
\noi and
 
\bea
\sigma(\bar\nu_ee\to\ell\bar\nu_\ell )= \frac{2G^2_\mu m_e E_\nu}{3\pi}
\left( 1+\frac{\alpha}{\pi} F_+\right) \nonumber \\
F_+ = -\frac23 \ell n \frac{2E_\nu}{m_e} -\frac16
\left(\pi^2-\frac{43}{4}\right) \label{eqthirtysix}
\eea
 
\noi Note, the $\ell n \frac{2m_e E_\nu}{m^2_\ell}$ terms cancel as
expected for total cross-sections.
 
For very high energy neutrinos, for example $E_\nu\simeq 40$ GeV, where
the above results are applicable, those corrections decrease the
cross-sections by 2.0\% and 1.8\% respectively. They must be included
in future high precision studies where $\nu_\mu e\to\mu^-\nu_e$ and/or
$\bar\nu_ee\to\mu^-\bar\nu_\mu$ is used to normalize the flux at say
0.1\% or better.
 
The radiative corrections to the neutral current and mixed reactions in
eqs.~(\ref{eqfour}) and (\ref{eqfive}) are somewhat more involved. One
problem is that the loop corrections are $q^2$ dependent. However, their
variation is not very significant for the range $0<-q^2 <2m_e E_\nu$, in
the
case of realistic $E_\nu$; so, we will approximate them by an average
$-q^2$ value. For a more thorough discussion see ref.~\cite{saran}.
 
Radiative corrections to the total NC cross-sections in
eqs.~(\ref{eqeighteen}) and (\ref{eqnineteen}) are obtained in three
steps \cite{saran}. First, $G_\mu$ is replaced by $\rho G_\mu$
 
\bea
G_\mu &\to& \rho G_\mu \nonumber \\
\rho &=& 1+\frac{\alpha}{4\pi} \left[\frac{3}{4s^4} \ell n c^2 -
\frac{7}{4s^2} + \frac{2}{c^2s^2} \left(\frac{19}{8} -\frac72 s^2 +
3s^4\right) \right. \label{eqthirtyseven} \\
&& \left. +\frac34 \frac{\xi}{s^2} \left[ \frac{\ell n
(c^2/\xi)}{c^2-\xi} +\frac{1}{c^2} \frac{\ell n \xi}{1-\xi} \right] +
\frac{3}{4s^2} \frac{m^2_t}{m^2_W} \right] \nonumber
\eea
 
\noi where $s^2\equiv \sintthw(m_Z)_{\overms}$, $c^2\equiv
\cos^2\theta_W(m_Z)_{\overms}$, $\xi= m^2_H/m^2_Z$ and $m_t\simeq 175$
GeV is the top quark mass. For $s^2=0.231$ and a Higgs mass, $m_H=130$
GeV, one finds
 
\be
\rho \simeq 1.013 \label{eqthirtyeight}
\ee
 
\noi which on its own would increase the NC cross-sections by 2.6\%.
 
The second effect of radiative corrections is to replace $\sintthw$ in
the tree level $\eps_-$ and $\eps_+$ by $\sintthw(q^2)$ which when
expressed in terms of the $\overline{MS}$ definition is given by (for
$\nnbar_\ell e$ scattering)
 
\be
\sintthw(q^2) = \kappa_\ell(q^2) \sintthw(m_Z)_{\overms}
\label{eqthirtynine}
\ee
 
\noi where
 
\bea
\kappa_\ell(q^2) &=& 1-\frac{\alpha}{2\pi s^2} \Biggl\{
2\sum_f(T_{3f}Q_f - 2s^2Q^2_f) J_f (q^2) -2 R_\ell
(q^2)\Biggr.\nonumber \\
&&\Biggl.+\frac{c^2}{3} +\frac12+\frac{1}{c^2} \left(
\frac{19}{8}-\frac{17}{4}s^2+3s^4\right) - \left(\frac72
c^2+\frac{1}{12}\right) \ell n c^2 \Biggr\} \label{eqforty}
\eea
 
\noi where
 
\bea
J_f (q^2) &=& \int^1_0 dx \; x(1-x) \ell n \frac{m^2_f-q^2
x(1-x)}{m^2_Z}  \qquad\qquad\qquad\qquad\qquad (41a) \nonumber \\
R_\ell(q^2) &=& \int^1_0 dx \; x (1-x) \ell n \frac{m^2_\ell -q^2
x(1-x)}{m^2_W} \qquad\qquad\qquad\qquad\qquad (41b) \nonumber
\eea
 
\noi The sum in eq.~(\ref{eqforty}) is over all fermions, with $T_{3f}
= \pm1/2$ and $Q_f={}$electric charge. For $q^2=0$, the quark
contributions have been evaluated using $e^+e^-\to{}$hadrons data in a
dispersion relation, one finds
 
\bea
\kappa_\mu(q^2=0) &=& 0.9970
\qquad\qquad\qquad\qquad\qquad\qquad\qquad\qquad\qquad\qquad (42a)
\nonumber \\
\kappa_\tau(q^2=0) &=& 1.0064
\qquad\qquad\qquad\qquad\qquad\qquad\qquad\qquad\qquad\qquad
(42b) \nonumber
\eea
 
\addtocounter{equation}{2}
 
\noi The difference between those two values can be viewed as a measure
of the different $\nu_\mu$ and $\nu_\tau$ charge radii \cite{degrassi}
(or anapole
moments for Majorana neutrinos). For $\langle q^2\rangle\simeq -0.02$
GeV$^2$, relevant for $E_\nu\simeq40$ GeV, the $\kappa$ in eq.~(42) are
reduced by about 0.001. Overall, the effect of the radiative correction
in eq.~(42a) is to decrease the $\nnbar_\mu e$ cross-sections by about
1\%.
 
The final source of radiative corrections comes from QED, including
bremsstrahlung. Those corrections are basically the same as in
$\nu_\ell e\to \ell \nu_e$ and $\bar\nu  e\to \ell \bar\nu_\ell$ for
left-left and left-right amplitudes, but with $m_\ell\to m_e$. In total
one finds \cite{saran}
 
\bea
\sigma(\nu_\ell e \to \nu_\ell e) = \frac{\rho^2 G^2_\mu m_e E_\nu}{2\pi}
&& \left[ (1-2\kappa_\ell (\bar q^2)s^2)^2 \left(1+\frac{\alpha}{\pi}
F_-\right) \right. \nonumber \\
&& \left. \quad +\frac13 (-2\kappa_\ell(\bar q^2)s^2)^2
\left(1+\frac{\alpha}{\pi} F_+\right) \right] \label{eqfortythree}
\eea
 
\noi where $F_-$ and $F_+$ are given in eqs.~(\ref{eqthirtyfive}) and
(\ref{eqthirtysix}) and $\bar q^2$ represents an average $q^2$. For
$\sigma (\bar\nu_\ell e\to\bar\nu_\ell e)$, one simply interchanges
$1-2\kappa_\ell s^2$ and $-2\kappa_\ell s^2$. Overall, the radiative
corrections tend to cancel and result in $\cal{O}$(1\%) shifts in the
cross-sections.
 
In the case of $\nnbar_ee\to \nnbar_ee$ cross-sections, the same
procedure as above applies except $G_\mu\to\rho G_\mu$ for the NC
amplitude while $G_\mu\to G_\mu$ for the CC amplitude. Also,
$\kappa_e(0)$ is smaller than $\kappa_\mu(0)$ by $\frac{\alpha}{3\pi
s^2} \ell n \frac{m_e}{m_\mu} \simeq - 0.0179$, because of the
significantly larger $\nu_e$ charge radius \cite{degrassi}.  That effect
is more
sensitive to $q^2$. In total, one finds \cite{saran}
 
\bea
\sigma(\nu_\e e \to \nu_\e e) = \frac{\rho^2 G^2_\mu m_e E_\nu}{2\pi}
&& \left[ (1-\frac{2}{\rho}-2\kappa_\e (\bar q^2)s^2)^2
\left(1+\frac{\alpha}{\pi}
F_-\right) \right. \nonumber \\
&& \left. \quad +\frac13 (-2\kappa_\e(\bar q^2)s^2)^2
\left(1+\frac{\alpha}{\pi} F_+\right) \right] \label{eqfortyfour}
\eea
 
\noi and $\sigma(\bar\nu_ee\to\bar\nu_ee)$ is obtained by interchanging
$1-\frac{2}{\rho} -2\kappa_e (\bar q^2)s^2$ and $-2\kappa_e(\bar
q^2)s^2$. The effect of radiative corrections on $\nnbar_ee$
cross-sections is overall more significant than $\nnbar_\mu e$.
 
If future measurements of $\nnbar e$ cross sections aim for $\pm0.1\%$
precision (or even better), the above radiative corrections must be
applied with care. Even some leading 2 loop effects should probably be
included. But why push those measurements to such extreme precision?
Some motivative, the search for ``New Physics'', will be given in
section~4.
 
\section{``New Physics'' Effects}
Very precise measurements of $\nnbar e$ scattering can in principle
test the SM at its quantum loop level and probe for ``New Physics''
effects. $G_\mu$ is extremely well determined via muon decay (see
eq(\ref{eqsix})) and $\sintthw(m_Z)_{\overms}$ has been determined to
better than $\pm0.1\%$
 
\be
\sintthw(m_Z)_{\overms} = 0.2312\pm0.0002 \label{eqfortyfive}
\ee
 
\noi Future efforts could further improve both of those quantities by
as much as an order of magnitude. Employing those values and the
explicit electroweak radiative corrections outlined in section~3 leads
to very precise predictions. For example, one finds the following SM
predictions
 
\moveleft.75in\vbox{\bea
\frac{d\sigma(\nu_\mu e\to\nu_\mu e)^{\rm SM}}{dy} &=& 0.2995 \left[
1+\frac{\alpha}{\pi} f_-(y) +0.7243 (1-y)^2
\left(1+\frac{\alpha}{\pi}f_+(y)\right) \right] \sigma(E_\nu)  (46a)
\nonumber
\\
\frac{d\sigma(\bar\nu_\mu e\to\bar\nu_\mu e)^{\rm SM}}{dy} &=& 0.2169
\left[
1+\frac{\alpha}{\pi} f_-(y) + 1.380 (1-y)^2
\left(1+\frac{\alpha}{\pi}f_+(y)\right) \right] \sigma(E_\nu) (46b)
\nonumber
\\
\frac{d\sigma(\nu_e e\to\nu_e e)^{\rm SM}}{dy} &=& 2.087 \left[
1+\frac{\alpha}{\pi} f_-(y) + 0.1003 (1-y)^2
\left(1+\frac{\alpha}{\pi}f_+(y)\right) \right] \sigma(E_\nu)  (46c)
\nonumber
\\
\frac{d\sigma(\bar\nu_e e\to\bar\nu_e e)^{\rm SM}}{dy} &=& 0.2093 \left[
1+\frac{\alpha}{\pi} f_-(y) + 9.969 (1-y)^2
\left(1+\frac{\alpha}{\pi}f_+(y)\right) \right] \sigma(E_\nu)  (46d)
\nonumber \\
&& \sigma(E_\nu) \equiv \frac{G^2_\mu m_e E_\nu}{2\pi}
\quad\qquad\qquad\qquad\qquad\qquad\qquad\qquad\quad
(46e) \nonumber
\eea}
 
\addtocounter{equation}{1}
 
\noi where the functions $f_-(y)$ and $f_+(y)$ were presented in
eqs.~(\ref{eqthirtytwo}) and (\ref{eqthirtythree}) (here $m_l$ is replaced by
 $m_e$). Those functions give
rise to large corrections to the final state electron spectrum see
ref.~\cite{saran}; however, they integrate only to about an overall 2\%
decrease.
Note, that we have given predictions to about 0.1\% to emphasize their
precision; but the numerical coefficients will actually change somewhat
for better specified experimental conditions (e.g.\ $q^2$ and $E_\nu$).
 
If measured $\nnbar e$ scattering cross-sections disagree with the SM
predictions in eq.~(46), it would be evidence for ``New Physics'' in
$\nnbar e$ scattering or $G_\mu$ and $\sintthw(m_Z)_{\overms}$
determinations. Some examples of ``New Physics'' that could cause such
deviations will be discussed in this section. To illustrate the
potential sensitivity of $\nnbar e$ scattering, we will sometimes
assume a future uncertainty of $\pm0.1$--0.5\% in those cross-sections
is achievable. Such a goal is very challenging. Its attainability will be
further discussed in section~6.
 
\subsection{Neutrino Dipole Moments}
We now believe that neutrinos have small masses and large mixing with
one another. If they are Dirac particles, they will have, albeit tiny,
magnetic dipole moments (and even smaller electric dipole moments). In
terms of electron Bohr magnetons, $e/2m_e$, one finds \cite{marcsix} the
SM prediction
 
\bea
\buildrel \rightharpoonup \over \mu_{\nu_i} &=& \frac{e}{m_e}
\kappa_{\nu_i} \buildrel \rightharpoonup \over S \nonumber \\
\kappa_{\nu_i} &=& \frac34 \frac{G_\mu m_e m_{\nu_i}}{\sqrt{2} \pi^2}
\simeq 3\times 10^{-19} (m_{\nu_i}/ eV) \label{eqfortyseven}
\eea
 
\noi Both Dirac and Majorana neutrinos can have transition moments that
link distinct mass eigenstates, giving rise to $\nu_2\to\nu_1+\gamma$,
for example.
 
Since $m_{\nu_i}$ are expected to be $<0.05$ eV, neutrino dipole
moments appear to be unobservable in the SM\null. However, in some
left-right symmetric models or extended Higgs models, it is possible to
have much larger dipole moments. It is, therefore, of interest to ask
what direct experimental bounds can be placed on neutrino dipole
moments (magnetic, electric, or transition), independent of theory?
 
Astrophysics considerations give the best constraints on
$\kappa\frac{e}{2m_e}$
 
\be
\kappa \le 10^{-12} \qquad (\rm Astrophysics) \label{eqfortyeight}
\ee
 
\noi but they depend on assumptions. It is useful to also obtain bounds
from neutrino scattering experiments, since they are direct.
 
The existence of any neutrino dipole moment (magnetic, electric or
transition) of magnitude $\kappa e/2m_e$ will increase $\nnbar e$
cross-sections by \cite{marc,kyul}
 
\be
\frac{\Delta d\sigma(\nu e)}{dy} = |\kappa|^2 \frac{\pi\alpha^2}{m^2_e}
\left(\frac{1}{y} -1\right) \label{eqfortynine}
\ee
 
\noi A larger than expected cross-section, particularly one exhibiting
a departure following the distinctive $1/y$ dependence of
eq.~(\ref{eqfortynine}) could be taken as evidence for a non-vanishing
$\kappa$. Consistency of current $\bar\nu_ee$ and $\nu_\mu e$
cross-sections with SM expectations gives the bounds \cite{marc}
 
\bea
|\kappa_{\nu_e}| &<& 4\times 10^{-10} \nonumber \\
|\kappa_{\nu_\mu}| &<& 10^{-9} \label{eqfifty}
\eea
 
\noi Those bounds could be translated to mass eigenstates by including
explicit mixing, but we do not carry out that exercise here.
 
It is quite difficult to do much better than the bounds in
eq~(\ref{eqfifty}) because of the $|\kappa|^2$ factor in
eq.~(\ref{eqfortynine}). Low energy neutrino beams are favored along
with $y$ dependence studies. For total cross-sections (with $y_{\rm
min}=0.01$), eq.~(\ref{eqfortynine}) leads to a fractional change in
$\sigma (\nu_\mu e\to\nu_\mu e)$ by a factor
 
\be
1+ 0.6\left|\frac{\kappa}{10^{-9}}\right|^2 \frac{1{\rm~GeV}}{E_\nu}
\label{eqfiftyone}
\ee
 
\noi Reaching $\kappa\simeq 10^{-10}$ sensitivity would require 0.006
GeV/$E_\nu$ precision. For high energy neutrinos $\gtsim 10$ GeV, that
might be difficult but perhaps not impossible. In the case of lower
energies, such as neutrinos from stopped pion decays,
e.g.\ $E_\nu\sim0.03$--1 GeV, it seems more straightforward. In fact,
at lower energies; such as neutrinos from stopped pion decays, one might
aim for a few $\times10^{-11}$ sensitivity,
particularly if the $y$ dependence is measured. Ultimately reaching
$|\kappa|\simeq 10^{-11}$ is extremely challenging, but provides a
worthwhile goal.  We should add that evidence for any neutrino dipole
moment significantly larger than the tiny SM predictions for such
quantities (e.g.\ eq(\ref{eqfortyseven})) would have important
implications, particularly for supernova physics \cite{lim}. Of course,
it would
also inspire theoretical explanation.
 
\subsection{Extra $\zp$ Bosons}
Many extensions of the SM predict the existence of additional neutral
$\zp$ bosons \cite{robin,marcseven}. They occur in SO(10), $E_6$ and
other grand unified
theories as well as in some superstring models. Evidence for their
existence would have profound implications for ``New Physics''.
 
Such bosons would lead to additional 4 fermion operators at low
energies whereas they would have little if any effect on $Z$ pole
properties. Currently, direct searches for $\zp$ bosons at the Tevatron
$p\bar p$ collider reach $\sim 350$--700 GeV \cite{marcfive,marcseven},
depending on couplings.
Low energy experiments such as atomic parity violation, neutrino
scattering, polarized $e^-e^-$ scattering etc.\ probe similar mass
scales. The LHC will push those searches to the multi-TeV region.
Future $e^+e^-$ colliders could indirectly probe even higher mass
scales via interference effects in cross sections and asymmetries.
Here, to roughly illustrate the sensitivity of $\nnbar$ scattering to
$\zp$ bosons, we consider the concrete example of the SO(10) $Z_\chi$
boson. The couplings for that case are fully specified and give rise to
an additional $\nnbar e$ (flavor independent) amplitude
 
\moveleft.55in\vbox{\be
M_{Z_\chi}=-i\frac{G_\mu}{\sqrt{2}} \frac34 \sintthw \bar u_\nu
\gamma^\alpha (1-\gamma_5) u_\nu \left[ \bar u_e\gamma_\alpha
(1-\gamma_5) u_e +\frac13 \bar u_e \gamma_\alpha (1+\gamma_5) u_e
\right] \frac{m^2_Z}{m^2_{Z_\chi}} \label{eqfiftytwo}
\ee}
 
\noi In that way, the $\eps_-$ and $\eps_+$ of eq.~(\ref{eqfourteen})
effectively become (using $s^2 \equiv \sintthw$)
 
\bea
\eps_- &\to& \frac12 - s^2 -\frac34 s^2 \frac{m^2_Z}{m^2_{Z_\chi}}
\nonumber \\
\eps_+ &\to& -s^2-\frac14 s^2\frac{m^2_Z}{m^2_{Z_\chi}}
\label{eqfiftythree}
\eea
 
\noi for pure neutral current scattering. One then finds for
$s^2=0.2312$ and $m^2_Z/m^2_{Z_\chi} << 1$
 
\moveleft.25in\vbox{\bea
\frac{d\sigma(\nu_\mu e\to\nu_\mu e)}{dy} &=& \frac{d\sigma(\nu_\mu
e\to\nu_\mu e)^{\rm SM}}{dy} + \sigma (E_\nu) (-0.37+0.11(1-y)^2)
\frac{m^2_Z}{m^2_{Z_\chi}}   (54a) \nonumber \\
\frac{d\sigma(\bar\nu_\mu e\to\bar\nu_\mu e)}{dy} &=&
\frac{d\sigma(\bar\nu_\mu
e\to\bar\nu_\mu e)^{\rm SM}}{dy} + \sigma (E_\nu) (0.11-0.37(1-y)^2)
\frac{m^2_Z}{m^2_{Z_\chi}} \;   (54b) \nonumber \\
\frac{d\sigma(\nu_e e\to\nu_e e)}{dy} &=& \frac{d\sigma(\nu_e
e\to\nu_e e)^{\rm SM}}{dy} + \sigma (E_\nu) (1.0+0.11 (1-y)^2)
\frac{m^2_Z}{m^2_{Z_\chi}} \quad  (54c)\nonumber \\
\frac{d\sigma(\bar\nu_e e\to\bar\nu_e e)}{dy} &=& \frac{d\sigma(\bar\nu_e
e\to\bar\nu_e e)^{\rm SM}}{dy} + \sigma (E_\nu) (0.11+1.0(1-y)^2)
\frac{m^2_Z}{m^2_{Z_\chi}} \quad  (54d) \nonumber
\eea}
 
\addtocounter{equation}{1}
 
\noi Comparison with eq.~(46) indicates $Z_\chi$ has the largest impact
on $\nu_\mu e$ scattering. Indeed, its integrated effect on
$\bar\nu_\mu e$ scattering is nearly zero. Of course, other $\zp$
models, extra dimensions, compositeness etc.\ can have quite different
influences.
 
A $\pm0.1\%$ measurement of $\sigma(\nu_\mu e\to\nu_\mu e)$ would probe
$m_\chi$ of 2--3 TeV\null. That appears to be a good benchmark for a
next generation experiment. Such precision might be possible at a
neutrino factory where high statistics are possible. However, if the
neutrinos stem from $\mu^-$ decays, one actually measures a spectrum 
flux weighted combination
of $d\sigma (\nu_\mu e)/dy + d\sigma (\bar\nu_ee)/dy$ at such a
facility. The weighting of each will depend on the muon polarization
and the region of $y$ explored. Therefore; good electron energy resolution
will be useful. Note that the simple integrated sum of eqs. (54a) and (54d)
showes little  $Z_\chi$ sensitivity.
 
At more conventional neutrino sources, i.e.\ horn focused beams or
stopped pions at an intense neutron spallation source, $\pm 0.5-1\%$
measurements of $\sigma(\nu_\mu e\to\nu_\mu e)$ are more realistic
expectations. At that level, $m_{Z_\chi}$ of order 800 - 1100 GeV would be
explored. That is similar to the capability of current precision
measurements in atomic parity violation, polarized $e^-e^-$ scattering
etc.
 
\subsection{Dynamical Symmetry Breaking}
Ideas such as technicolor provide interesting alternatives to the
elementary Higgs scalar mechanism. In those dynamical scenarios new
fermion condensates break SU(2)$_L \times{}$U(1)$_Y\to{}$U(1)$_{em}$. The
basic premise
of dynamical symmetry breaking is very appealing, but no simple
phenomenologically viable model currently exists.
 
A rather generic prediction of dynamical models is the existence of new
quantum loop effects due to effectively heavy fermions in gauge boson
self energies. A nice formalism for studying such loop effects is the
$S$, $T$ and $U$ parametrization \cite{peskin}, which is sometimes
expanded to
include additional $V$, $W$, $X$ and $Y$ corrections \cite{maks} due to
loop
changes between $|q^2|=0$ and $m^2_Z$. Here, we will not review that
formalism. Instead, we focus on $S$ and $T$, the most interesting of
those parameters, and show how non-zero values for those quantities
would affect $\nnbar e$ scattering. We should note that generically one
expects $S\simeq \frac{2}{3\pi}\simeq0.2$ for a very heavy fourth
generation of
fermions and similar or larger effects in dynamical symmetry breaking
scenarios. Also, currently precision electroweak measurements already
provide the constraints (from global fits) \cite{marcnine} (for
$m_H\simeq300$ GeV)
 
\bea
S &=& -0.11\pm0.11 \nonumber \\
 T &=& -0.07\pm0.13 \label{eqfiftyfive}
\eea
 
\noi Those bounds are already quite restrictive. The constraint on $S$
severely limits dynamical symmetry breaking scenarios and seems to
imply that a heavy fourth generation of fermions is very unlikely.
Continuing to search for non-vanishing $S$ and $T$ is strongly
warranted, but individual next generation experiments should aim for
$\pm0.1$ or better $S$ and $T$ sensitivity to be competitive.
 
Within the framework where $\alpha$, $G_\mu$ and $m_Z$ are fixed by
their experimental values, the SM plus new heavy fermion loops predicts
modifications in $\sigma (\nnbar e)$ resulting from the shifts
\cite{marceight}
 
\bea
\rho = \rho^{\rm SM} (1+0.0078 T) \nonumber \\
\sintthw(m_Z)_{\overms} = 0.2312+0.00365S-0.0026T \label{eqfiftysix}
\eea
 
\noi One sees that roughly speaking, measuring individual
$\sigma(\nnbar_\mu e)$ and $\sintthw(m_Z)_{\overms}$ to about
$\pm0.1\%$ would determine $S$ and $T$ to better that $\pm0.1$. More
specifically, we consider the shift in $\nnbar e$ cross-sections due to
$S$ and $T$
 
\be
\Delta \frac{d\sigma}{dy} = \frac{d\sigma}{dy} -
\frac{d\sigma^{\rm SM}}{dy} \label{eqfiftyseven}
\ee
 
\noi One finds
 
\moveleft.7in\vbox{\bea
\frac{\Delta d\sigma(\nu_\mu e\to\nu_\mu e)}{dy} &\simeq&
[(0.01T-0.008S) +(1-y)^2 (-0.0015T+0.007S)] \sigma(E_\nu)  (58a)
\nonumber \\
\frac{\Delta d\sigma(\bar\nu_\mu e\to\bar\nu_\mu e)}{dy} &\simeq&
[(-0.0015T+0.007S) +(1-y)^2 (0.01T-0.008S)] \sigma(E_\nu)  (58b)
\nonumber \\
\frac{\Delta d\sigma(\nu_e e\to\nu_e e)}{dy} &\simeq&
[(-0.027T+0.02S) +(1-y)^2 (-0.0015T+0.007S)] \sigma(E_\nu)  (58c)
\nonumber \\
\frac{\Delta d\sigma(\bar\nu_e e\to\bar\nu_e e)}{dy} &\simeq&
[(-0.0015T+0.007S) +(1-y)^2 (-0.027T+0.02S)] \sigma(E_\nu)  (58d)
\nonumber
\eea}
 
\addtocounter{equation}{1}

An interesting possibility  at a neutrino factory is to use a combination of
$\sigma(\nu_\mu e\to\nu_\mu e)  + \sigma(\bar\nu_ee\to \bar\nu_ee)$
with each weighted by spectral flux functions. That combination can be
normalized using the inverse muon decay reactions $\sigma(\nu_\mu e\to
\mu^-\nu_e) + \sigma(\bar\nu_ee\to \mu^-\nu_\mu)$ which are $S$ and $T$
independent in the above formalism. Note that $R_4$ obtained using that 
normalization when fully integrated over $y$ has reduced $S$ and $T$ 
dependence.  Detailed studies of neutrino factory
capabilities \cite{mang}  suggest that $\pm0.1\%$ determinations of
$\sintthw(m_Z)_{\overms}$ may be possible. At that level, $S$ and $T$
sensitivity will be about $\pm0.1$. So, we conclude that $\nnbar
e$ scattering measurements can potentially be competitive next generation
probes of $S$ and $T$, but experiments must be capable of roughly
$\pm0.1\%$ sensitivity for $\sigma(\nu e)$ and
$\sintthw(m_Z)_{\overms}$. Also, good final state electron energy
resolution will be very useful in unfolding the $y$ dependence. The
quantity $R_3$ in eq.~(\ref{eqtwentyeight}) will be a particularly
sensitive probe if it can be measured with small systematic
uncertainties.
 
\section{Neutrino Indices of Refraction}
Neutrino propagation through matter can be sensitive to neutrino
indices of refraction. Those indices result from forward scattering
amplitudes for neutrino scattering with the constituents of matter,
including electrons \cite{wolf}. That makes this topic appropriate for an
article
on neutrino-electron scattering.
 
The amplitude for low energy $\nu_\ell$ scattering off a fermion $f$,
where $\ell$ now stands for $e$, $\mu$ or $\tau$ is given by
 
\be
M(\nu_\ell f+\nu_\ell f) = -i\frac{G_\mu}{\sqrt{2}} \bar\nu_\ell
\gamma^\alpha (1-\gamma_5)\nu_\ell \bar f\gamma_\alpha (C^V_{\nu_\ell
f} + C^A_{\nu_\ell f} \gamma_5) f \label{eqfiftynine}
\ee
 
\noi For an unpolarized medium, the coherent forward scattering of
neutrinos with momentum $p_\nu$ can be described by an index of
refraction $n_{\nu_\ell}$ given by \cite{marcten}
 
\be
p_\nu (n_{\nu_\ell}-1) = - \sqrt{2} G_\mu \sum_{f=e,u,d} C^V_{\nu_\ell f}
N_f \label{eqsixty}
\ee
 
\noi where $N_f$ is the fermion number density and at tree level in the
SM
 
\bea
C^V_{\nu_\ell f} &=& T_{3f} -2Q_f\sintthw \qquad f\ne\ell \nonumber \\
C^V_{\nu_\ell \ell} &=& 1+T_{3\ell} -2Q_\ell\sintthw \label{eqsixtyone}
\eea
 
\noi with $T_{3f} = \pm 1/2$. The difference between $\nnbar_e e$ and
$\nnbar_\mu e$, $\nnbar_\tau e$ scattering cross-sections due to CC
interactions in the former gives rise to a difference in the indices of
refraction
 
\be
-p_\nu (n_{\nu_e} - n_{\nu_\mu}) = -p_\nu (n_{\nu_e} -n_{\nu_\tau}) =
\sqrt{2} G_{\mu} N_e \label{eqsixtytwo}
\ee
 
\noi That difference can significantly impact neutrino oscillations in
matter \cite{wolf}.
 
To gain insight into the origin of eq.~(\ref{eqsixtytwo}), it is useful
to consider $-iM(\nu_\ell f\to\nu_\ell f)$ as an effective Lagrangian.
Averaging that Lagrangian over the background matter medium, one finds
\cite{marcten}
 
\be
\langle C^V_{\nu_\ell f} \bar f \gamma_0 f\rangle = C^V_{\nu_\ell f}
N_f \label{eqsixtythree}
\ee
 
\noi whereas (for an unpolarized medium) all other currents average to
zero. So, the medium can be interpreted as providing an external
potential
 
\be
V=\sqrt{2} G_\mu \sum_f C^V_{\nu_\ell f} N_f \label{eqsixtyfour}
\ee
 
\noi experienced by propagating neutrinos. With such a potential
 
\be
i \frac{d}{dt} \to i \frac{d}{dt} - V \label{sixtyfive}
\ee
 
\noi in the equation of motion (for antineutrinos $V=-V$). Although the
potential is small
 
\be
|V| \sim 4\times10^{-14} {\rm~eV~} (N_f/6\times10^{23}cm^{-3}),
\label{eqsixtysix}
\ee
 
\noi it can have truly remarkable consequences when it interferes with
other very small effects such as neutrino energy differences
$m^2_i-m^2_j/2E_\nu$, neutrino dipole moment precession in magnetic
fields \cite{lim,akh}, neutrino decay in matter etc. We will not review
those
interesting topics. Instead, we conclude this discussion by reviewing
radiative corrections to indices of refraction.
 
It was shown in ref.~\cite{botella} that the radiative corrections to
 
\be
p_\nu (n_{\nu_e}-n_{\nu_\mu}) = -\sqrt{2} G_\mu N_e
\label{eqsixtyseven}
\ee
 
\noi are negligible, $\cal{O}$$(\alpha m^2_\mu/m^2_W)$. They do, however,
give rise to an interesting one loop induced
$p_\nu(n_{\nu_\tau}-n_{\nu_\mu})$ of order $\alpha m^2_\tau/m^2_W$. For
an unpolarized medium and with $N_n=N_p=N_e$, one finds \cite{botella}
 
\be
n_{\nu_\tau}-n_{\nu_\mu} \simeq 5\times10^{-5} (n_{\nu_e}-n_{\nu_\mu})
\label{eqsixtyeight}
\ee
 
\noi Although small, that loop induced difference can be of some
importance in supernova studies where extremely high densities are
possible.
 
Of course, if ``New Physics'' exists which differentiates $\nu_e$,
$\nu_\mu$ and $\nu_\tau$ interactions at the tree level, it can have
dramatic effects on neutrino oscillations in matter. Future studies of
neutrino oscillations over very long terrestrial baselines will provide
interesting probes of such interactions.
 
\section{Outlook}
Currently, there are no precision $\nnbar e$ scattering studies
underway.  The only use of that set of reactions is for solar neutrino
flux measurements by Super K\null. However, the discovery of
oscillations has invigorated neutrino physics. Future efforts to
measure neutrino mixing and mass parameters with high precision and
search for new phenomena will demand high intensity neutrino sources.
One can envision using those facilities to carry out precision studies
of other neutrino properties at short baselines where neutrino
oscillations may not be operational, but other phenomena can be
explored. Here, we will not comment on neutrino oscillations or the
many other
very interesting phenomena that can be studied with new intense
neutrino sources, although they will provide the primary motivation for
such facilities. Instead, we conclude this article with a brief
perspective on the utility of low, medium and high energy intense
neutrino facilities for studying $\nnbar e$ scattering.
 
\subsection{Low Energy}
Very intense spallation neutron sources are copious sources of pions.
The decay chain $\pi^+\to\mu^+\nu_\mu$, $\mu^+\to e^+\nu_e\bar\nu_\mu$
for stopped $\pi^+$ would provide equal fluxes of low energy $\nu_\mu$,
$\nu_e$ and $\bar\nu_\mu$ with very well predicted energy spectra. The
latter  two are time separated from the $\nu_\mu$ by the muon lifetime.
One can, therefore, contemplate the measurement of $R_2$ in
eq.~(23) with high precision. An unfulfilled LAMPF
proposal \cite{white} would have measured that ratio to $\pm1.7\%$. One
can imagine
using low energy neutrinos from future (several megawatt) spallation
neutron facilities to push that goal to about $\pm0.5\%$ (perhaps
further if systematics can be controlled). At that level $m_{\zp}$ of
$\cal{O}$(1 TeV) would be probed and a sensitivity of $\pm0.2$ in $S$
and $T$ could be achieved. Because of the small neutrino energy, bounds
on $\nu_e$ and $\nu_\mu$ dipole moments of about $2\times10^{-11}
e/m_e$ would also be possible.
 
The goal of 0.5\% for $R_2$ is difficult but not unrealistic. The
neutron spallation sources will exist and the detector requirements are
not that demanding. Such studies are not so costly and well worth the
effort.
 
\subsection{Superbeams}
Using conventional horn focused pions from intense proton sources (1
megawatt or more), $\nu_\mu$ or $\bar\nu_\mu$ beams of high intensity
(sometimes called superbeams) are possible. Their average energy (if spawned by protons of energy
$\sim 28$ GeV) will be $\sim 1$--2 GeV\null. They could be used to
statistically measure $\sigma(\nu_\mu e\to \nu_\mu e)$ and
$\sigma(\bar\nu_\mu e\to \bar\nu_\mu e)$ to $\pm0.5\%$ or better. A
systematic limitation will be flux spectrum normalization. Achieving
$\pm0.5\%$ or better will be very challenging.
 
If a $\pm0.5\%$ determination of  $\sigma(\nu_\mu e\to \nu_\mu e)$ and
$\sigma(\bar\nu_\mu e\to \bar\nu_\mu e)$ is possible, it will explore
$m_{\zp}$ of $\cal{O}$(1 TeV) and $S$, $T\sim\pm0.2$. To do well on
neutrino dipole moments, it must map out the $y$ dependence of those
cross-sections. A detailed study is necessary before one can
confidently assess the capabilities of (conventional) superbeams for
doing more than very long baseline neutrino oscillations, an area where
they are extremely well motivated \cite{marceleven, parsaEpac02,diwan}.
 
\subsection{Neutrino Factories}
High energy neutrinos ($\sim40$ GeV) from muon storage rings are
particularly interesting for studying cross-sections (at
short-baselines). Their flux scales as $E^2_\mu$ and cross-sections
grow with $E_\nu$ making high statistics very straightforward. An
attractive case for measuring structure functions, QCD effects, CKM
elements etc.\ has been made \cite{mang}.
 
Even for $\nnbar e\to \nnbar e$ scattering, such a facility is
potentially very powerful. With integrated luminosities of $\sim
10^{47}$cm$^{-2}$ and $\nnbar e$ cross-sections of $\sim
10^{-39}-10^{-40}$cm$^{2}$, one can envision statistics of
$10^7$--$10^8$ events at such a facility in each of the $\nnbar e$
scattering modes \cite{forte}. The inverse muon decay cross-sections
$\sigma(\nu_\mu
e\to\mu^-\nu_e) + \sigma(\bar\nu_e\to\mu\bar\nu_\mu)$ can be used (at
least for $\mu^-$ decay neutrinos) to normalize the $\nu_\mu$ and
$\bar\nu_e$ spectra to $\pm0.1\%$ or better, a rather unique
capability. Neutrinos originating from $\mu^+$ decays will be harder to
normalize, but their relative cross-sections will carry much information.
 
If $\pm0.1\%$ precision on the various $\nnbar e\to \nnbar e$
cross-sections can be achieved \cite{mang}, it will represent a major
advance.
Within the SM, that will allow $\Delta\sintthw\simeq\pm0.1\%$. In terms
of the ``New Physics'' capabilities $m_{\zp}$ of several TeV will be
explored and $\Delta S$, $\Delta T$ approaching $\pm0.05$ may be
possible. These are impressive capabilities, particularly since they
represent a small part of the intended full program. Of course, the
overall cost of a high energy neutrino factory is prohibitive. It will
be interesting to see if the strong physics case is enough to justify
the facility.
 
Neutrino physics is difficult, but well worth pursuing. Its pursuit
will not only push forward physics frontiers, but will demand
technological innovation. Intellectual curiosity is truly the Mother of
invention and catalyst for the ascent of science.


\bigskip
\begin{thebibliography}{99}
 
 
 
\leftline{\bf References}
\bigskip
 
\bibitem{one} Gargamelle collaboration, F. Hasert {\it et al}., Phys.\
Lett.\ {\bf B46}, 121 (1973).  
 
\bibitem{two} see J. Panman in ``Precision Tests of the Standard
Electroweak Model'', ed.\ P. Langacker (World Scientific 1995) p.~504
and references therein.
 
\bibitem{three} Super $K$ Collaboration, S. Fukuda {\it et al}., Phys.\
Lett.\ {\bf B539}, 179 (2002) 
 
\bibitem{marc} see W. Marciano and Z. Parsa, Ann.\ Rev.\ Nucl.\ Part.\
Sci.\ {\bf36}, 171 (1986) and references therein.  
 
\bibitem{mang} see M. Mangano {\it et al}.\ hep-ph/0105155; C. Albright
{\it et al}.\ hep-ex/0008064; I. Bigi {\it et al}.\ hep-ph/0106177; Z.
Parsa, AIP Conference Proc.\ {\bf 549}, 781 (2002).
Those studies present a physics case for short baseline physics at a
Neutrino Factory.  
 
\bibitem{saran} S. Sarantakos, A. Sirlin and W. Marciano, Nucl.\ Phys.\
{\bf B217}, 84 (1983).  
 
\bibitem{marctwo} See W. Marciano in  ``Fifty Years of Electroweak
Physics'', ed.\ L. Chatterjee, J. Phys.\ {\bf G29}, 23 (2003). 
 
\bibitem{thooft} G. 'tHooft, Phys.\ Lett.\ {\bf B37}, 195 (1971)  
 
\bibitem{white} D.H. White {\it et al}., LAMPF Cherenkov Detector
Proposal No.\ 1015, 1986 (unpublished).   
 
\bibitem{pass} J. Bahcall, M. Kamionkowski and A. Sirlin, Phys.\ Rev.\
{\bf D51}, 6146 (1995); M. Passera, Phys.\ Rev.\ {\bf D64}, 113002
(2001).  
 
\bibitem{marcthree} W. Marciano and A. Sirlin, Phys.\ Rev.\ {\bf D22},
2695 (1980); Nucl.\ Phys.\ {\bf B189}, 442 (1981).  
 
\bibitem{marcfour} W. Marciano and A. Sirlin, Phys.\ Rev.\ Lett.\ {\bf
46}, 163 (1981).  
 
\bibitem{sirlin} A. Sirlin, Phys.\ Rev.\ {\bf D22}, 971 (1980).  
 
\bibitem{marcfive} W. Marciano and A. Sirlin, Phys.\ Rev.\ {\bf D27},
552 (1983); W. Marciano in ``Spin Structure in High Energy Processes'',
Proceedings of the 21st SLAC Summer Institute, Stanford, California
1993, edited by L. De Porcel and C. Dunwoodie (SLAC Report No.\ 444,
Stanford, 1994).  
 
\bibitem{degrassi} G. Degrassi, A. Sirlin and W. Marciano, Phys.\ Rev.\
{\bf D39}, 287 (1989); W. Marciano and A. Sirlin in ``Neutrino
Physics'' edited by K. Winter, (Cambridge 1991, 2000) p.~186.   
 
\bibitem{marcsix} W. Marciano and A. Sanda, Phys.\ Lett.\ {\bf B67},
303 (1977); B. Lee and R. Shrock, Phys.\ Rev.\ {\bf D16}, 1444 (1977);
M.A.B. Beg, W. Marciano and M. Ruderman, Phys.\ Rev.\ {\bf D17}, 1395
(1978); K. Fujikawa and R. Shrock, Phys.\ Rev.\ Lett.\ {\bf 55}, 963
(1980).  
 
\bibitem{kayser} B. Kayser, Phys.\ Rev.\ {\bf D26}, 1662 (1982).  
 
\bibitem{kyul} A. Kyuldjiev, Nucl.\ Phys.\ {\bf B243}, 387 (1984).
 
\bibitem{lim} C.S. Lim and W. Marciano, Phys.\ Rev.\ {\bf D37}, 1368
(1988).  
 
\bibitem{robin} R. Robinett and J.L. Rosner, Phys.\ Rev.\ {\bf D25},
3036 (1982); N. Deshpande and D. Iskander, Nucl.\ Phys.\ {\bf B167}, 223
(1980); P. Langacker, R. Robinett and J.L. Rosner, Phys.\ Rev.\ {\bf
D30}, 1470 (1984).  
 
\bibitem{marcseven} W. Marciano in ``Precision Tests of the Standard
Model'' edited by P. Langacker (World Scientific 1995) p.~170.   
 
\bibitem{marceight} W. Marciano and J. Rosner, Phys.\ Rev.\ Lett.\ {\bf
65}, 2963 (1990).   
 
\bibitem{czar}  A. Czarnecki and W. Marciano, Phys.\ Rev.\ {\bf D53},
1066 (1996).  
 
\bibitem{peskin} M. Peskin and T. Takeuchi, Phys.\ Rev.\ Lett.\
{\bf65}, 964 (1990); Phys.\ Rev.\ {\bf D46}, 381 (1992).  
 
\bibitem{maks} I. Maksymyk, C. Burgess and D. London, Phys.\ Rev.\ {\bf
D50}, 529 (1994); A. Kundu and P. Roy, hep-ph/9411225, Int.\ J. Mod.\
Phys.\ {\bf A12}, 1511 (1997).  
 
\bibitem{marcnine} W. Marciano, hep-ph/9903451, Phys.\ Rev.\ {\bf D60},
093006 (1999); J. Erler and P. Langacker, hep-ph/9809352; PDG K.
Hagiwara {\it et al}., Phys.\ Rev.\ {\bf D66}, 1 (2002).  
 
\bibitem{wolf} L. Wolfenstein, Phys.\ Rev.\ {\bf D17}, 2369 (1978);
{\bf 20}, 2634 (1979); P. Mikheyev and A. Smirnov, Nuovo Cimento {\bf
9C}, 17 (1986); P. Langacker, J. Leveille and J. Sheiman, Phys.\ Rev.\
{\bf D27}, 1228 (1983).   
 
\bibitem{marcten}  W. Marciano in Proceedings of the 3rd Conf.\ on
Interactions between Particle and Nucl.\ Phys., ed.\ G. Bunce 1988,
p.~193.  
 
\bibitem{akh} E. Akhmedov, Phys.\ Lett.\ {\bf B213}, 64 (1988).  
 
\bibitem{botella} F. Botella, C.-S. Lim and W. Marciano, Phys.\ Rev.\
{\bf D35}, 896 (1987).  
 
\bibitem{marceleven} W. Marciano, hep-ph/0108181.  
 
\bibitem{parsaEpac02} Z. Parsa in Proceedings of EPAC, Paris, France,  
 ed.\ J. Pool and Leonid Rivkin (2002), 1036. 

\bibitem{diwan} M. Diwan {\it et al}., hep-ph/0303081, Phys.\ Rev.\
{\bf D68}, 012002 (2003); Z. Parsa, in Proceedings of the 8th Conf. on
Interactions of  Particle and Nucl. Phys. (2003).  
 
\bibitem{forte} S. Forte, hep-ph/0109219.   
 
\end{thebibliography}
\end{document}